# Majority and Minority Voted Redundancy for Safety-Critical Applications

P. Balasubramanian, D.L. Maskell  
School of Computer Science and Engineering  
Nanyang Technological University  
Singapore 639798

N.E. Mastorakis  
Department of Industrial Engineering  
Technical University of Sofia  
Sofia 1000, Bulgaria

*Abstract*—A new majority and minority voted redundancy (MMR) scheme is proposed that can provide the same degree of fault tolerance as N-modular redundancy (NMR) but with fewer function units and a less sophisticated voting logic. Example NMR and MMR circuits were implemented using a 32/28nm CMOS process and compared. The results show that MMR circuits dissipate less power, occupy less area, and encounter less critical path delay than the corresponding NMR circuits while providing the same degree of fault tolerance. Hence the MMR is a promising alternative to the NMR to efficiently implement high levels of redundancy in safety-critical applications.

*Keywords*— Redundancy, Fault tolerance, Low power, ASIC, Combinational logic, Standard cells, CMOS

## I. Introduction

Safety-critical applications such as aerospace systems, nuclear power plants, electricity transmission and distribution facilities, banking and financial systems, industrial control and automation, and other sensitive industry applications usually incorporate redundancy in their physical designs to embed a specific degree of fault tolerance to successfully overcome arbitrary function units' faults or failures [1], with the function unit being any circuit or system. In this context, the N-modular redundancy (NMR) scheme has been widely used. However, the drawbacks with NMR are the exacerbated increases in the number of function units and their corresponding design metrics, and weight and cost, to achieve higher levels of redundancy. The NMR, shown in Fig. 1, employs N identical function units and requires the correct operation of at least (N+1)/2 function units, which represents a Boolean majority. The NMR voter performs majority voting on the outputs of all the N function units to determine the NMR output.

The 3MR, which is the 3-tuple version of the NMR, uses 3 identical function units and requires the correct operation of at least 2 function units. On the other hand, the 5MR, 7MR and 9MR, which are the respective 5-tuple, 7-tuple and 9-tuple versions of the NMR require 5, 7 and 9 function units, and mandate the correct operation of at least 3, 4 and 5 function units respectively. The issue with NMR is that to achieve enhanced fault tolerance the number of function units to be employed would disproportionately increase. For example, when the fault tolerance is to be increased by unity from 2 (i.e., 5MR) to 3 (i.e., 7MR) two additional function units should be introduced in the NMR scheme. Besides, the corresponding majority voter would considerably increase in size although NMR voters can be optimally realized using multiplexers [2].

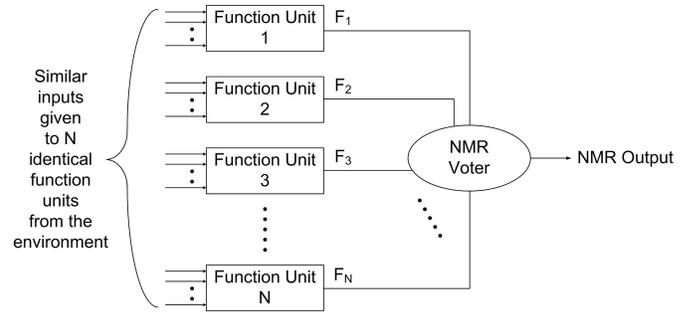

Fig. 1. NMR architecture

In modern electronic designs, multiple faults or failures are known to occur owing to continued miniaturizations of the transistor dimensions and the associated parametric variations, and due to the adverse impact of radiation and other phenomena on small device geometries [3 – 8]. To cope with multiple faults or failures, higher levels of redundancy such as 5MR, 7MR and 9MR are suggested to be used selectively in the sensitive portions of a circuit or system [9]. However, the drawback with NMR is the requirement of more function units which would have a direct bearing on the design metrics, and weight and cost. To achieve enhanced fault tolerance with fewer function units and weight and cost, this paper proposes a new redundancy scheme called the majority and minority voted redundancy (MMR) that requires relatively fewer function units than the NMR and can achieve significant reductions in the design metrics and weight and cost.

The remainder of this paper is organized as follows. Section 2 presents the proposed MMR architecture, describes its operation, and discusses the reliability of the MMR versus the NMR. Section 3 gives the design metrics estimated for example MMR and NMR circuits, which were implemented using a 32/28nm CMOS process. Finally, Section 4 provides the conclusions.

## II. MMR – Architecture, Operation and Reliability

The generic architecture of the MMR scheme is shown in Fig. 2. The MMR employs K identical function units and hence

This work is supported by the Singapore Ministry of Education (MoE) Academic Research Fund Tier 2 under grant MOE2017-T2-1-002 and MoE Tier 1 under grant RG132/16.

it is also referred to as 'K-MMR'. The K identical function units are grouped into two clusters viz. the majority cluster and the minority cluster, which are depicted using the blue and pink boxes in Fig. 2. The outputs of these clusters i.e., $F_1$ to $F_K$ are combined using the MMR voter, portrayed within the orange box, to produce the MMR output. The majority cluster consists of function units 1, 2 and 3, and the minority cluster comprises the remaining (K–3) function units. Note that the number of function units in the majority cluster is kept a constant (i.e., 3), and the Boolean majority condition is only imposed on the three function units comprising the majority cluster. This contrasts with the NMR where the Boolean majority condition is imposed on all the N function units. Function unit(s) can be added to the minority cluster of the MMR to enhance its fault tolerance as desired. The two operational conditions imposed on the MMR architecture to guarantee its correct operation are: i) at least 2 out of the 3 function units in the majority cluster should operate correctly, and ii) at least 1 out of the (K–3) function units in the minority cluster should operate correctly. Hence, the Boolean majority and minority conditions are imposed on the majority and minority clusters respectively.

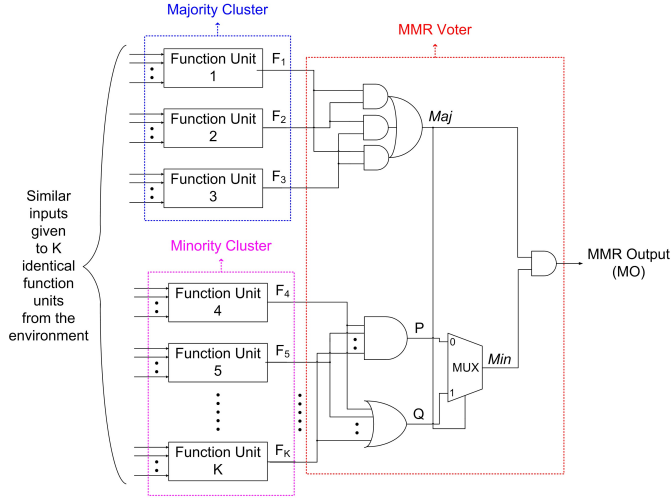

Fig. 2. Proposed MMR architecture

The MMR voter, shown in Fig. 2, comprises a 3-input majority gate [10, 11], also called the AO222 complex gate, that performs majority voting on the outputs of the majority cluster and produces *Maj*. Notably, *Maj* serves as the reference output for the MMR. *Min* cannot be the reference output for the MMR since it would give rise to ambiguity because of the Boolean minority condition. For example, assuming $F_4$ is 0 and $F_5$ up to $F_K$ are 1 in the minority cluster, the Boolean minority condition would interpret both 0 and 1 as corresponding to the Boolean minority. This is because at least one of $F_4$ to $F_K$ is 0, and at least one of $F_4$ to $F_K$ is 1 simultaneously. On the other hand, if for example $F_1 = F_2 = 0$ and $F_3 = 1$ in the majority cluster, according to the Boolean majority condition, *Maj* = 0, and there is no ambiguity.

The MMR voter also comprises a (K–3)-input AND gate and a similar size OR gate. These gates can be arbitrarily decomposed for minimum area or delay and hence they are synthesizable. The outputs of the AND and OR gates are given to a 2:1 multiplexer (MUX), which forms a part of the MMR voter, and whose select input is *Maj*. *Min* represents the output of the minority cluster, and the logical conjunction of *Maj* and *Min* using a 2-input AND gate yields the primary MMR output, MO, as shown in Fig. 2.

The operation of the MMR scheme is illustrated through Table I by highlighting three scenarios. Table I captures all the possible output combinations with respect to the majority cluster and representative subsets of the possible output combinations with respect to the minority cluster. In Table I, the notations used for the minority cluster imply the following: i) '$F_4 - F_K$' given by '0 – 0' or '1 – 1' implies $F_4$ up to $F_K$ are all binary 0 or 1 respectively, ii) '$F_4 - F_K$' given by '0 – 1' implies $F_4$ is 0, and $F_5$ up to $F_K$ may be 1, and iii) '$F_4 - F_K$' given by '1 – 0' implies $F_4$ is 1, and $F_5$ up to $F_K$ may be 0.

Referring to Fig. 2 and Table I, and considering scenario 1, if *Maj* = 0, and if $F_4$ up to $F_K$ are all 0s, P = Q = 0, *Min* = 0, and hence MO = 0. Alternatively, if *Maj* = 1, and if $F_4$ up to $F_K$ is 1, P = Q = 1, and *Min* = 1, and hence MO = 1. Given scenario 2, when *Maj* = 0, and if at least one of $F_4$ up to $F_K$ is 0, P = 0 and Q = 1. Since *Maj* = 0, P is selected and its value is forwarded to *Min*, and *Min* = 0. Subsequently, MO would correctly assume 0. With respect to scenario 3, given *Maj* = 1, and if at least one of $F_4$ up to $F_K$ is 1, P = 0 and Q = 1. Since *Maj* = 1, Q is selected and its value is forwarded to *Min*, and *Min* = 1. Therefore, MO would correctly assume 1.

TABLE I. ILLUSTRATING THE OPERATION OF MMR

| Majority Cluster | | | Minority Cluster | | MMR Internal Voter Outputs | | MMR Output |
|---|---|---|---|---|---|---|---|
| $F_1$ | $F_2$ | $F_3$ | $F_4$ | – $F_K$ | *Maj* | *Min* | MO |
| *Scenario 1: Majority and minority clusters are perfect* | | | | | | | |
| 0 | 0 | 0 | 0 | – 0 | 0 | 0 | 0 |
| 1 | 1 | 1 | 1 | – 1 | 1 | 1 | 1 |
| *Scenario 2: Majority and minority clusters are imperfect, and majority cluster outputs* **0** | | | | | | | |
| 0 | 0 | 1 | 0 | – 1 | 0 | 0 | 0 |
| 0 | 1 | 0 | 0 | – 1 | 0 | 0 | 0 |
| 1 | 0 | 0 | 0 | – 1 | 0 | 0 | 0 |
| *Scenario 3: Majority and minority clusters are imperfect, and majority cluster outputs* **1** | | | | | | | |
| 1 | 1 | 0 | 1 | – 0 | 1 | 1 | 1 |
| 1 | 0 | 1 | 1 | – 0 | 1 | 1 | 1 |
| 0 | 1 | 1 | 1 | – 0 | 1 | 1 | 1 |

In the K-MMR it is required that at least two function units belonging to the majority cluster and one function unit belonging to the minority cluster should operate correctly. Thus, the fault tolerance of the K-MMR is given by (K–3). The fault tolerance of the NMR is specified by (N–1)/2. Given this, the 5-MMR, 6-MMR and 7-MMR can tolerate the faulty or the failure states of maximum of 2, 3 and 4 function units respectively. As a result, the 5-MMR, 6-MMR and 7-MMR form the corresponding redundant counterparts of the 5MR, 7MR and 9MR based on fault tolerance. The system or circuit reliability equations of the 5-MMR, 6-MMR and 7-MMR are given by (1) to (3). These equations are derived based on the following assumptions: i) K identical function units are used in the K-MMR, and the reliability of the K function units are considered to be equal; let a function unit reliability be represented by R, where R is a function of time $t$ i.e., R($t$), and ii) the perfect behaviour of MMR voters.

$$R^{\text{5-MMR}} = 6R^3(1-R)^2 + 5R^4(1-R) + R^5 \quad (1)$$

$$R^{\text{6-MMR}} = 9R^3(1-R)^3 + 12R^4(1-R)^2 + 6R^5(1-R) + R^6 \quad (2)$$

$$R^{\text{7-MMR}} = 12R^3(1-R)^4 + 22R^4(1-R)^3 + 18R^5(1-R)^2 + 7R^6(1-R) + R^7 \quad (3)$$

In (1), the first term on the right side represents the condition when 2 out of the 3 function units in the majority cluster, and 1 out of the 2 function units in the minority cluster operate correctly. The second term specifies the condition of 2 function units in the majority cluster and 2 function units in the minority cluster operating correctly or the correct operations of all the 3 function units in the majority cluster and only 1 function unit in the minority cluster. The last term signifies the condition of all the 5 function units operating correctly.

In (2), the first term on the right side specifies the condition of 2 function units in the majority cluster and 1 function unit in the minority cluster operating correctly. The second term indicates the correct operation of 2 function units in the majority cluster and 1 function unit in the minority cluster or the correct operations of all the 3 function units in the majority cluster and 1 function unit in the minority cluster. The third term signifies the condition of any 2 function units in the majority cluster and all the 3 function units in the minority cluster operating correctly or the correct operations of all the 3 function units in the majority cluster and any 2 function units in the minority cluster. The last term specifies the condition of all the function units maintaining the correct operation.

Referring to (3), the first term on the right side specifies the correct operation of 2 out of the 3 function units in the majority cluster and 1 function unit in the minority cluster. The second term indicates the correct operation of 2 function units in the majority cluster and 2 function units in the minority cluster or the correct operations of all the 3 function units in the majority cluster and 1 function unit in the minority cluster. The third term signifies the correct operation of all the 3 function units in the majority cluster and 2 function units in the minority cluster or the correct operations of any 2 function units belonging to the majority cluster and any 3 function units belonging to the minority cluster. The fourth term reflects the correct operation of all the 3 function units present in the majority cluster and the 3 function units present in the minority cluster or the correct operations of any 2 function units present in the majority cluster and all the 4 function units present in the minority cluster. The last term indicates the correct operation of all the function units present in both the clusters.

Fig. 3 shows a plot of the reliabilities of K-MMR and corresponding NMR implementations versus their function unit reliability. The reliabilities of the respective voters are not considered to simplify the calculations. It can be seen that the system/circuit reliabilities of the former are slightly less than the latter. For a function unit reliability ranging from 0.9 to 0.99, which is typical of a safety-critical application [12], the 5-MMR, 6-MMR and 7-MMR report 1.21%, 1.06% and 1.08% less reliability than the 5MR, 7MR and 9MR respectively.

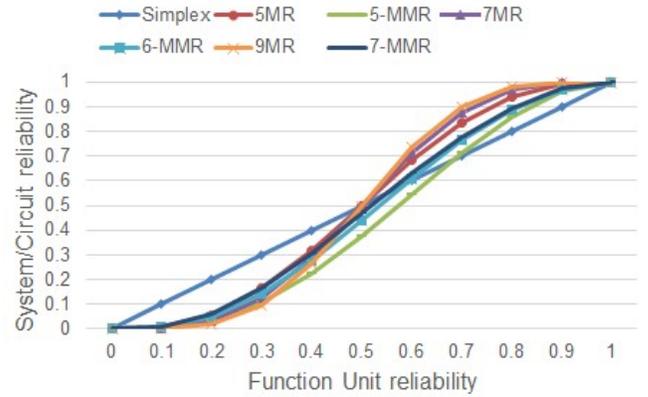

Fig. 3. Comparison of the reliabilities of NMR and MMR

III. MMR AND NMR – EXAMPLE IMPLEMENTATIONS

Example NMR and MMR circuits were physically realized using a 4-bit ripple carry adder (RCA) and a 4×4 binary array multiplier (BAM) for the function units, separately. A 32/28nm bulk CMOS standard digital cell library [13] was used for the physical implementations. The functional simulations were performed by supplying all the distinct input vectors identically to all the function units at time intervals of 2.5ns (400MHz) to verify the implementations. The switching activities captured through the functional simulations were used to estimate the average power dissipation. The design metrics such as average power dissipation, critical path delay, and area were estimated using Synopsys tools and are given in Table II. Overall, the MMR circuits exhibit less critical path delay, occupy less area, and dissipate less power than the corresponding NMR circuits.

TABLE II. DESIGN METRICS OF NMR AND CORRESPONDING MMR CIRCUITS

| Function Unit | Redundancy Type | Delay (ns) | Area (µm²) | Power (µW) |
|---|---|---|---|---|
| 4-bit RCA | 5MR | 0.64 | 156.30 | 56.14 |
| | 5-MMR (proposed) | 0.67 | 152.49 | 52.74 |
| | 7MR | 0.77 | 296.08 | 92.68 |
| | 6-MMR (proposed) | 0.67 | 172.82 | 61.03 |
| | 9MR | 0.88 | 479.06 | 144.2 |
| | 7-MMR (proposed) | 0.67 | 198.23 | 70.74 |
| 4×4 BAM | 5MR | 0.98 | 529.64 | 120.7 |
| | 5-MMR (proposed) | 1.01 | 523.54 | 116.4 |
| | 7MR | 1.12 | 865.11 | 191.2 |
| | 6-MMR (proposed) | 1.01 | 611.98 | 137.0 |
| | 9MR | 1.23 | 1269.7 | 278.5 |
| | 7-MMR (proposed) | 1.01 | 708.55 | 159.3 |

The critical path delays of NMR circuits increase with increases in the level of redundancy. This is because the size of the NMR voters increases with increases in the level of redundancy, which is accompanied by associated increases in the corresponding logic depth. Increases in the logic depth of NMR voter circuits cause increases in the critical path delay. On the contrary, the critical path delays of MMR circuits are less compared to the corresponding NMR circuits.

As the level of redundancy increases, the design metrics of NMR circuits increase substantially while the design metrics of MMR circuits increase only gradually. This is mainly because for every extra function unit introduced in the minority cluster of the MMR its fault tolerance increases by unity while the number of function units in the majority cluster is kept as constant i.e., 3. In comparison, the NMR requires the inclusion of 2 extra function units to improve its fault tolerance by unity.

An NMR voter considerably increases in size with an increase in the level of redundancy while the increase in the corresponding MMR voter size is just gradual. The 5MR, 7MR and 9MR voters occupy respective silicon areas of 13.47μm$^2$, 34.31μm$^2$ and 63.79μm$^2$. In contrast, the 5-MMR, 6-MMR and 7-MMR voters occupy reduced areas of 12.71μm$^2$, 13.26μm$^2$ and 14.74μm$^2$ of silicon respectively. Fig. 4 shows the distribution of power dissipation between the function units and voters of NMR and MMR circuits. The power dissipations of function units and voters increase significantly in the case of the NMR with increases in the level of redundancy, and the power dissipations of function units and voters increase only gradually in the case of the proposed MMR.

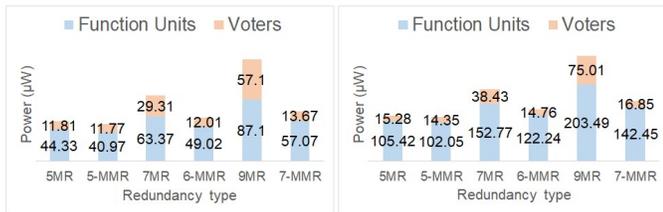

Fig. 4. Split-up of average power dissipation between the function units and voters of NMR and counterpart MMR circuits with: (a) 4-bit RCA used as the function unit, and (b) 4×4 BAM used as the function unit

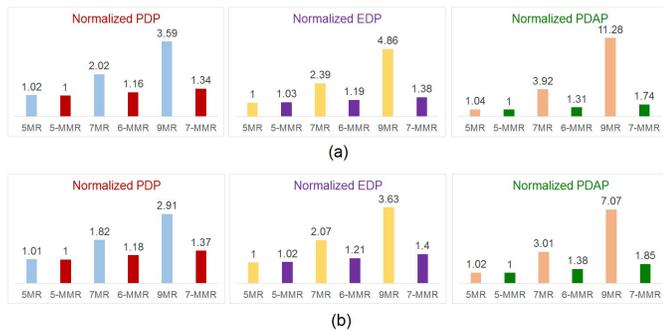

Fig. 5. Normalized FOMs of NMR and MMR circuits with: (a) 4-bit RCA used as the function unit, and (b) 4×4 BAM used as the function unit

Fig. 5 depicts three qualitative figure-of-merits (FOMs) viz. power-delay product (PDP), energy-delay product (EDP), and power-delay-area product (PDAP) for the NMR and MMR circuits. Fig. 5a shows the normalized FOMs based on using the 4-bit RCA for the function units, and Fig. 5b shows the normalized FOMs based on using the 4×4 BAM for the function units. Since power, delay, and area parameters are desirable to be minimized, the least values of PDP, EDP and PDAP indicate the best design. Fig. 5 shows that, overall, MMR circuits report better FOMs i.e. reduced PDP, EDP and PDAP than their counterpart NMR circuits for various levels of redundancy. Also, the increases in PDP, EDP and PDAP are found to be significant in the case of NMR circuits but they are noted to be only gradual in the case of MMR circuits for increases in the redundancy.

IV. CONCLUSION

This paper presented a new redundancy scheme for fault-tolerant design i.e., MMR (also referred to as K-MMR). MMR requires fewer function units and a less sophisticated voting logic based on Boolean majority and minority conditions, which could facilitate reduced design metrics, and weight and cost than the conventional NMR to achieve the same degree of fault tolerance. Hence, the MMR is a promising alternative to the NMR to effectively implement high levels of redundancy (selectively) in safety-critical applications.